\documentclass[runningheads]{llncs}
\usepackage{llncsdoc}

\usepackage{amssymb}
\usepackage{graphicx}
\usepackage{textcomp}
\usepackage{url}

\usepackage{url}
\newcommand{\keywords}[1]{\par\addvspace\baselineskip
\noindent\keywordname\enspace\ignorespaces#1}

\begin{document}

\title{A Personalized Tag-Based Recommendation in Social Web Systems}

\titlerunning{A personalized tag-based recommendation in social web systems}

\author{Frederico Durao and Peter Dolog}
\authorrunning{Frederico Durao and Peter Dolog}

\institute{IWIS --- Intelligent Web and Information Systems,\\ Aalborg University, Computer Science Department\\
Selma Lagerl{\"{o}}fs Vej 300, DK-9220 Aalborg-East, Denmark\\
\texttt{\{fred,dolog\}@cs.aau.dk}}

\maketitle

\begin{abstract}

Tagging activity has been recently identified as a potential source of knowledge about personal interests, preferences, goals, and other attributes known from user models. Tags themselves can be therefore used for finding personalized recommendations of items. In this paper, we present a tag-based recommender system which suggests similar Web pages based on the similarity of their tags from a Web 2.0 tagging application. The proposed approach extends the basic similarity calculus with external factors such as tag popularity, tag representativeness and the affinity between user and tag. In order to study and evaluate the recommender system, we have conducted an experiment involving 38 people from 12 countries using data from Del.icio.us, a social bookmarking web system on which users can share their personal bookmarks.

\end{abstract}

\section{Introduction}

Collaborative tagging systems have become increasingly popular for sharing and organizing Web resources, leading to a huge amount of user generated metadata. Tags in social bookmarking systems such as \textit{del.ici.ous}\footnote{http://delicious.com} are usually assigned to conceptualize, categorize, or sharing a resource on the Web so that users can be reminded of them later and find their bookmarks in an easy way. Invariably, tags represent some sort of affinity between user and a resource on the web. By tagging, users label resources on the Internet freely and subjectively, based on their sense of values \cite{tagging}. In this sense, tags from social bookmarking systems represent a potential mean for personalized recommendation because through them it is possible to identify individual and common interests between unknown users. Nevertheless, although huge amount of tag data is available, to compute an individual preference in order to perform efficient recommendation is still a challenging task. In this paper, we propose a tag-based recommender system which recommends bookmarks by calculating the similarity of their tags. The proposed approach besides basic similarity takes into account external factors such as \textit{tag popularity}, \textit{tag representativeness} and the \textit{affinity between user and tag}. We utilize a cosine similarity measure between tag vectors to calculate basic similarity of the pages. We measure tag popularity as a count of occurrences of a certain tag in the total amount of web pages. We utilize term frequency measure to compute tag representativeness for a certain web page. The tag affinity between a user and a tag is calculated as a count of how many times the user utilized the tag at different web pages. We propose a formula which considers all these factors in a normalized way and gives a ranking of web pages for particular user.

The goal of this study is to analyze whether tags can be utilized to generate personalized recommendations. This assumption can be assessed by running an experiment whereby users expresses their satisfaction about the received recommendations. Based on this, we conducted an experiment involving 38 people from 12 countries using data from del.icio.us to evaluate the efficiency of the proposed approach and social aspects such as the purposes behind the tagging activity. The contribution of the paper is therefore:
\begin{itemize}
\item{The proposed recommendation approach based on similarity, tag representativeness, popularity, and affinity; and}
\item{Findings from the evaluation which show that the approach performs well  in a non-controlled environment with people from different domains and intentions.}
\end{itemize}
The paper is organized as follows: In Section 2 we discuss related work. In Section 3 we introduce the motivation of this work. Section 4 describes the factors of similarity that we will analyze. Section 5 presents the experiment and the achieved results. Section 6 addresses a discussion about the findings from the experiment and Section 7 presents the conclusion and future works.

\section{Related Work} \label{s-work}

Tags have been recently studied in the context of recommender systems due to various reasons. \cite{Nakamoto2007} argues for a solution where tagging from social bookmarking provides a context for recommender systems in terms of context clues from tags as well as connectivity among users to improve the collaborative recommender system. Similar to our approach, \cite{Niwa} constructed a web recommender based on large amount of public bookmark data on Social Bookmarking system. For means of personalization, they utilize folksonomy tags to classify web pages and to express user's preferences. By clustering folksonomy tags, they can adjust the abstraction level of user's preferences to the appropriate level. In spite of the proximity with our study, the \cite{Niwa} experiment did not measure the efficiency of the recommendations in terms of user satisfaction what could have provided us a parameter for comparison. \cite{Pavan} extends a content based recommender system by deriving current and general personal interests of users from different tags according to different time intervals. However, unlike our approach, the similarity of the tags is given by of two Naive Bayes classifiers trained over different timeframes: one classifier predicts the user's current interest whereas the other classifier predicts the user's general interest in a bookmark. The two classifiers are trained with a subset of the bookmarks created by a user. The tags of each bookmark, converted into a "bag of words", are used as training features. The bookmarks are recommended in case both of the two classifiers predict a bookmark as interesting. The effectiveness of the recommendations however is totally dependent on the quality of the subset of bookmarks used for training the classifiers.

\cite{FiranNP07} shows the benefits of using tag based profiles for personalized recommendations of music on Last.fm. Similar understanding over the product items as subject of recommendations is considered as another factor in addition to the similar tags when personalizing recommendations given by a tag based collaborative recommender system in \cite{Zhao2008}. The purpose of tags vary as well as tagging itself may be influenced by different factors.  For example, \cite{SenLRCFOHR06} studies a model for tagging evolution based on community influence and personal tendency. It shows how 4 different options to display tags affect user's tagging behavior. \cite{BischoffFNP08} studies how the tags are used for search purposes. It confirms that the tags can represent different purpose such as topic, self reference, and so on and that the distribution of usage between the purposes vary across the domains. Other works such as \cite{Staab02} and \cite{Mika07} coined the term {\em emergent semantics} as the semantics which emerge in communities as social agreement on tag's meaning based on its more frequent usage instead of the contract given by ontologies from ontology engineering point of view. However, the approaches based on emergent semantics are characterized by the power law which gives a long tail of the tags of which semantics have not emerged yet. Therefore, \cite{CattutoBHS08} looks at grounding of the tag relatedness with a help of WordNet. 

In this paper we look at, how multiple factors such as similarity, tag popularity, tag affinity to a user and tag representativeness can be used together to achieve recommendations. We also wanted to see the personalized recommendations in an open context with users of different background.

\section{Running Example or Motivating Scenario}

Tags in social bookmarking systems allow users to express their preference by sharing their bookmarks. Tags are personalized piece of information which can be utilized to identify common interests between users. Compared to traditional collaborative rating \cite{colRating}, tags can reflect the user's preference to a given resource in a meaningful way \cite{tagging}. Based on these premises, we investigate the feasibility of using tags as one approach for the generation of \textit{personalized recommendations}. Along this article, the word \textit{resource} will be used as generic term to refer to document, video, image, text, files or any sort of asset which can be tagged and referenced by URI.

\begin{figure}
    \centering
    \includegraphics[width=3.0in]{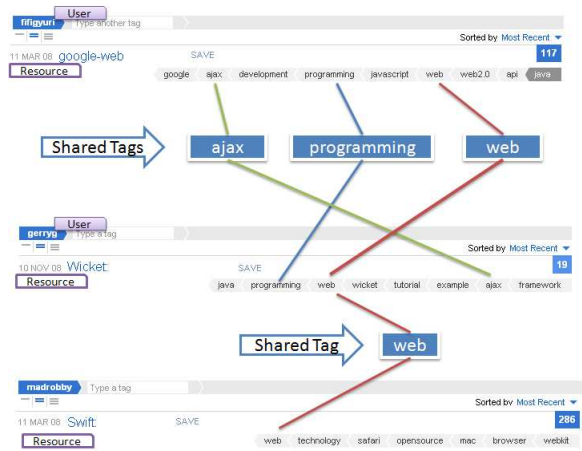}
    \caption{User x Bookmarks x Tags}
    \label{triple}
\end{figure}

Let us now look at a scenario which explains our approach and ideas behind it. According to Figure \ref{triple}, the resources \textbf{Google-Web} and \textbf{Wicket} share tags \textit{ajax},\textit{programming} and\textit{web} whereas resources \textit{Wicket} and \textit{Swift} only share tag \textbf{web}. Considering exclusively the similarity between tags, the resources \textbf{Google-Web} and \textbf{Wicket} have higher probability of being about the same content than \textbf{Wicket} and \textbf{Swift}. Based on this fact, their authors should be noticed about the existence of similar resources around. Furthermore, the notification could be prioritized (or emphasized) if the similar tags correspond to the most frequent tags of the authors or they are very representative for the resource they describe. This scenario was presented for illustrating how tag similarity can be computed for means of personalization. Although similarities can be found when the tags are syntactically identical, a number of pessimist scenarios may take place and must be considered such as: \textit{resources which have similar tags but incorrect spelling} - since tags are informal and free writing, no syntax control is assured. For instance, tags "programming" and "programing" looks the same except for the fact the second one is missing the letter \textit{m}; \textit{resources whose tags are syntactically different but similar semantically} - this is a case of synonymy and to overcome it some semantic assistance is needed either by use of domain ontologies or looking up for synonymies in dictionary. For instance, tags "work" and "labor" looks different but share the same meaning. The obstacle is that generic dictionary sometimes is not enough to provide the correct meaning of specific terms in a given \textit{context}; and \textit{resources which share same tags with different meanings} - this is the well known case of polysemy. For instance, the tag "windows" can be about the operational system or the house artifact.

\section{The Approach}

In order to generate personalized recommendations, we propose an extension method for the calculation of basic similarity between tags. We combine the cosine similarity calculus with other factors such as \textit{tag popularity}, \textit{tag representativeness} and\textit{ affinity user-tag} with the purpose of reordering the original raking in recommendation and generate personalized ones.

We define the document score as:

\noindent $Ds = \displaystyle\sum_{i=1}^n weight(Tag_i) * \displaystyle\sum_{i=1}^n representativness(Tag_i)$, where \textit{n} is the total number of existing tags in the repository.

We define the tag user affinity as:

\noindent $ Affinity_{(u, t)} = card\{r \in Documents \mid (u, t, r) \in R, R \subseteq U \times T \times D\} / card\{t \in T \mid (t, u) \in R_u, R_u \subseteq U \times T\}$, where \textit{t} is a particular tag, \textit{u} particular user, \textit{U} is a set of users, \textit{D} set of resources and \textit{T} set of tags.

Finally, similarity is computed as:

\noindent $Similarity_{(D_i, D_{ii})} = [Ds_{D_i} + Ds_{D_{ii}} * cosine\_similarity(T_{D_i},T_{D_ii})] * Affinity_{(u, t)}$,  where \textit{Ds} is the document score and \textit{T} is set of tags of a particular document.

Informally, each one of the factors in the above formulas is calculated as follows:
\begin{itemize}

\item {
\textbf{Cosine Similarity} --- Our tag similarity is a variant on the classical cosine similarity familiar from text mining and information retrieval \cite{cosineSimilairtyMetric} whereby two items are thought of as two vectors in the \textit{m} dimensional user-space. The similarity between them is measured by computing the cosine of the angle between these two vectors.} 

\item  {
\textbf{Tag Popularity} --- Also called \textit{tag weight}, is calculated as a count of occurrences of one tag per total of resources available. We rely on the fact that the most popular tags are like anchors to the most confident resources. As a consequence, it decreases the chance of dissatisfaction by the receivers of the recommendations.}

\item  {
\textbf{Tag Representativeness} --- It measures how much a tag can represent a document it belongs. It is believed that those tags which most appear in the document can better represent it. The \textit{tag representativeness} is measured by the \textit{term frequency}, a broad metric also used by the Information Retrieval community.}

\item  {
\textbf{Affinity between user and tag} - It measures how often a tag is used by a user. It is believed that the most frequent tags of a particular user can reveal his/her interests. This information is regarded as valuable information for personalization means. During the comparison of two resources, the similarity is boosted if one of the resources contains top tags of the Author from the other resources around.}
\end{itemize}

Further, we have set empirically that for one tag represent the user's preference, its frequency of use must be 70\% closer to the most frequent tag of the user. In the case on which there are no tags to satisfy this condition, it is assumed the user does not have a clear preference.

\section{Experimental Evaluation}
\label{sec:experiment}

In our evaluation, we opted to measure the degree of satisfaction of users about the received recommendations. The user's feedback will allow us to evaluate the quality of recommendations produced from our framework. Although, we recognize that \textit{precision and recall} are metrics which could be used to evaluate the effectiveness of the system, we believe that user's participation provides more precious feedback for means of personalization. In this sense, we invited users by sending a number of invites in various mailing lists from different natures, not only related to technology. We explained the purpose of the experiment and also we outlined easiness of the participation aim at attracting more users not related to technology. To our surprise, within less than 1 month 44 participants had accepted to participate voluntarily. In spite of 44 initial positive replies, only 38 participants joined until the end. Finally, we had 38 participants from 12 countries interested in many different subjects. Data for our experiment was collected from \textit{del.icio.us} in November 2008 comprising 5542 tags and 1143 bookmarks.

\subsubsection{Methodology.}
\label{sec:methodology}
We have created a \textit{del.icio.us} user account for each participant on which he/she was invited to add at least 10 bookmarks with minimally 3 tags each as suggested. Each participant received the top 5 most similar recommendations to their bookmarks based on the tags assigned to them. Then the participants were asked to select which items of the recommended set matched to their bookmarks. As soon as the participants finished their contribution, the overall results were shared with the participants as well as the reflections and findings.

\subsection{Expected Results}
\label{sec:expectedResults}
Considering that the experiment took place in non-controlled environment (as del.ici.ous is) with diverse audience (people from technology, health, education, biology, etc), we did not expected 100\% of acceptance of the recommendations. Some reasons for this are: i) \textit{diversity of culture and background} - Since the participants are from many different countries and have distinct backgrounds, it increases significantly the disparity between tags i)\textit {Syntax of tags} - As previously introduced, the tags assigned by the participants were not under any syntax control. Users could have written their tags in many different (and personalized) ways, for instance, the tag \textit{web2.0} can be also tagged as \textit{web20}, \textit{web2\_0} or \textit{web\_20} and iii) \textit{difficulty to identify user's preferences through the tags} - if users bookmark web resources of different domains (e.g. sports, education, engineering), hardly any some tags will predominate over others, which increases the difficulty of precisely indentifying user's main preferences.

Based on the reasons addressed previously, we consider the result as satisfactory if more than half of recommendations are accepted (or selected) by the participants. If 80\% of the recommendations are accepted, we claim the results as excellent (and unexpected), on the other hand, bellow of 50\%, we understand that the proposal has to be reviewed and improved with the findings achieved from the experiment analysis.

\subsection{Results from the experiment}
\label{sec:results}
\begin{figure}
    \centering
\includegraphics[width=4.0in]{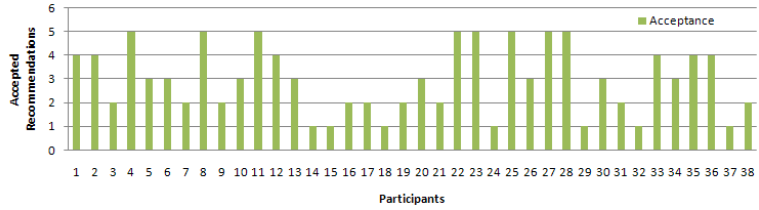}
    \caption{Amount of recommendations accepted by each participant.}
    \label{acceptance}
\end{figure}

Figure \ref{acceptance} provides a clear picture of how many items were accepted by each participant among the five recommended. For instance, the participant 9 accepted 2 recommendations amongst the 5 suggested in the set. 

The graph shows that at least one recommendation from each set recommended was accepted. Moreover, it shows that 8 participants were satisfied with the whole set of recommendations and 7 participants accepted only 1 item from whole the set which was recommended. Due to the graph distribution, it is possible to preliminary argue that the acceptance is well balanced. However, in order to evaluate the overall results more properly, we stipulated a threshold by calculating the \textit{arithmetic mean} of the acceptance, which was \textit{2.971}. In the following, we calculated the \textit{standard error of the mean} in order to verify the variance of the mean and consequently perform more concrete argumentation on top of the results obtained. The standard error is given by  $se=\frac{s}{\sqrt{n}}$, where \textit{s} is the sample standard deviation (i.e. the measure of the dispersion of the data set), \textit{n} is the size (number of observations) of the sample. More about the standard deviation can be found at \cite{standardDeviation}. Finally, the standard error of the mean obtained was \textit{0.23}, which allow us to judge our results based on the stipulated threshold (2.971) without significant variance.

\begin{figure}
    \centering
\includegraphics[width=2.5in]{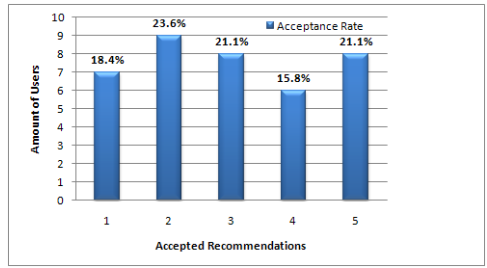}
    \caption{Frequency of recommendations accepted}
    \label{h1}
\end{figure}

Figure \ref{h1} shows the histogram of the accepted recommendations. It shows the frequency of the accepted items against the amount of participants. Figure \ref{h1} shows that 22 participants (or 58\% of all) accepted 3 or more recommendations (above the threshold) from the five that was suggested. However, 16 (or 42\% of all) participants accepted only 2 or 1 recommendations, below the stipulated threshold. Focusing only on the 16 participants who accepted between 2 or 1 items, 7 of them accepted only 1 recommendation from the whole set. This means that 7 participants together rejected 80\% of the recommendations that were sent to them (i.e. 35 sent and 28 rejected). In order to investigate this particular inconvenience, we analyzed the tags assigned to these rejected items. We figured out that although the recommendations had been generated correctly (considering the tag syntax), most of them was really out of context and far from the user's interest. We turn out with some findings: i) some tags did not show clearly any relatedness with the resource domain; ii) ambiguity problems, more particularly synonymy, when the tags share same syntax but are semantically different; and iii) impossibility of identifying user's preference due to the low number of tags of some users.

\begin{figure}
    \centering
\includegraphics[width=2.5in]{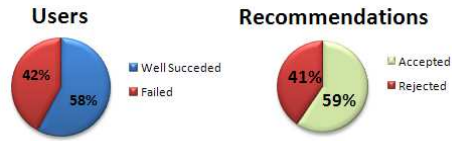}
    \caption{Final results}
    \label{finalResults}
\end{figure}

Figure \ref{finalResults} summarizes the overall results. As already pointed, the pie chart on the left shows that 58\% of the participants received a set of recommendations above the stipulated threshold while 42\% received a set of recommendations bellow of it. The pie chart on the right shows the overall rate of well succeeded recommendations in which 59\% of the whole recommendations was accepted and 41\% was rejected. In summary, the final result cannot be considered excellent but satisfactory, since 58\% of the overall set of recommendations was above the threshold and 59\% of the overall recommendations was accepted. In spite of achieving satisfactory results, it is not possible to affirm the proposed solution is ready for large usage. Improvements to overcome the ambiguity problems are needed and further experiments must be performed again in order to provide more insights about the system evolution. On the other hand, the results indicate the research is on the right track.

\section{Discussion}
\label{sec:discussion}

From the evaluation, we realized that the proposed recommender approach performs satisfactorily well even in a non-controlled environment with users from different domains and backgrounds. Based on the results, we understand the multifactor approach can be utilized to generate personalized recommendations. However, it is quite important to discuss the problems found from the unsuccessful recommendations. The ambiguity problems such as synonymy and polysemy can be attenuated by using WordNet dictionary since it has been employed for computing semantic similarity \cite{CattutoBHS08}; however, they are generic and do not cover particular meanings from specific domains. Focusing on the problem of (or lacking of) relatedness between tags and resource, we believe that a viable solution is to capture the purpose why users are tagging as studied by \cite{BischoffFNP08}. If the purpose is asked explicitly, then we have a usability problem, i.e. one additional step to simply assign a tag to a resource. On the other hand, to infer the purpose of a tag in a given resource relies on long observation about the user´s tagging activity. Moreover, if the inference is uncertain, a number of bad recommendations can be processed. Concerning the difficulty of identifying the user's preference using tags, we understand that the factor \textit{time} should be taken into account. The user's preference changes along the time and these changes can be reflected in the tags as well.

\section{Conclusion and Future Works}
\label{sec:discFinal}

This paper introduced a tag-based recommender system which generate personalized recommendations. The efficiency of the system rely on cosine similarity calculus with additional factors such as \textit{tag popularity}, \textit{tag representativeness} and \textit{affinity between user and tag}. A experiment involving 38 people from 12 countries using data from del.icio.us was conducted to evaluate the efficiency of the system for means of personalization.

The overall results showed that approximately 60\% of the recommendations succeeded and the proposed recommender system requires improvements. As a future work, we propose to perform semantic similarity to overcome ambiguity problems (as mentioned in the pessimist scenarios) and investigate the purpose of the tags when they are assigned to a resource. Finally, comparisons with other approaches must be addressed since the current evaluation methodology only assesses user's satisfaction using the specific algorithm.

\section{Acknowledgment}
\label{sec:Acknowledgment}
The research leading to these results is part of the project "KiWi - Knowledge
in a Wiki" and has received funding from the European Community's Seventh
Framework Programme (FP7/2007-2013) under grant agreement No. 211932.

\bibliographystyle{plain}
\bibliography{umap2009}
\end{document}